\documentstyle[aps,epsf]{revtex}
\newcommand{\be}{\begin{eqnarray}}
\newcommand{\ee}{\end{eqnarray}}

\newcommand{\real}{\mbox{{\rm I\hspace{-2truemm} R}}}

\begin{document}
\draft
%
%
%
\title{Black Hole Evaporation and Compact Extra Dimensions}
\author{Roberto Casadio$^a$ and Benjamin Harms$^b$}
\address{~}
\address{$^a\,$Dipartimento di Fisica, Universit\`a di Bologna,
and I.N.F.N., Sezione di Bologna,
via Irnerio 46, I-40126 Bologna, Italy}
%
\address{$^b\,$Department of Physics and Astronomy,
The University of Alabama,
Box 870324, Tuscaloosa, AL 35487-0324, USA}
\maketitle
\date{\today}
\begin{abstract}
We study the evaporation of black holes in space-times with
extra dimensions of size $L$.
We first obtain a description which interpolates between the
expected behaviors of very large and very small black holes
and then show that the luminosity is greatly damped when the
horizon shrinks towards $L$ from a larger value.
Analogously, black holes born with an initial size smaller than
$L$ are almost stable.
This effect is due to the dependence of both the Hawking
temperature and the grey-body factor of a black hole on the
dimensionality of space.
Although the picture of what happens when the horizon
becomes of size $L$ is still incomplete, we argue that there
occurs a (first order) phase transition, possibly signaled by
an outburst of energy which leaves a quasi-stable remnant.
\end{abstract}
\section{Introduction}
\setcounter{equation}{0}
\label{intro}
The current interest in the possibility that there exist large extra
dimensions \cite{arkani,RS} beyond the four dimensions of our every
day experience is based on two attractive features of this proposal.
First of all the hierarchy problem is by-passed because the radiative
stability of the weak scale in this proposal is due to the
identification of the ultraviolet cutoff of the theory with the
electroweak energy scale $\Lambda_{EW}$.
Ancillary assumptions such as supersymmetry are not required to achieve
radiative stability.
The other attraction of this proposal is that since the fundamental
scale of the theory is the electroweak energy scale, predictions drawn
from the theory such as deviations from the $1/r^2$ law of Newtonian
gravity can be experimentally tested in the near future.
In the extra-dimensions scenario all of the interactions, gravity as
well gauge interactions, become unified at the electroweak scale.
This means that if the model is viable, particle accelerators such as
the LHC, the VLHC and the NLC will be able to uncover the features of
quantum gravity as well as the mechanism of electroweak symmetry
breaking.
\par
The large-extra-dimension scenario also has significant implications
for processes involving strong gravitational fields, such as the
bending of light near a massive object and the decay of black holes.
The latter phenomenon has been described within the context of the
microcanonical ensemble (see Refs.~\cite{r1,mfd}) in four space-time
dimensions.
Our starting point is the idea that black holes are (excitations of)
extended objects ($p$-branes), a gas of which satisfies the bootstrap
condition.
This yields a picture in which a black hole and the emitted particles
are of the same nature and an improved law of black hole decay which is
consistent with unitarity (energy conservation).
This idea has been bolstered by recent results in fundamental
string theory, where it is now accepted that extended D$p$-branes are
a basic ingredient \cite{polchinski}.
States of such objects were constructed which represent black holes
\cite{vafa} and corroborate \cite{maldacena} the old idea that the
area of the horizon is a measure of the quantum degeneracy of the black
hole \cite{st}.
However, this latter approach works mostly for very tiny black holes
and suffers from the same shortcoming that the determination of the
space-time geometry during the evaporation is missing.
\par
Embedding a black hole in a space-time of higher dimensionality would
seem, from the string theory point of view, to be a natural thing to do.
In models with extra spatial dimensions the four dimensional spacetime
is viewed as a D3-brane embedded in a bulk spacetime of dimension $4+d$.
In the most recent manifestation of string theory, matter is described
by open strings whose endpoints are Dirichlet branes.
Therefore in the large-extra-dimension scenario matter fields at low
energy (in the present case this means at energies less than
$\Lambda_{EW}$) are confined to live on the D3-brane.
In string theory gravity is described by closed strings and hence it
can propagate in the bulk.
Black holes in $4+d$ extra dimensions have been studied in both compact
\cite{argyres} and infinitely extended \cite{chamblin} extra dimensions
(see also \cite{emparan} and references therein).
\par
In Ref.~\cite{ch} we studied the evaporation of black holes in
space-times with extra dimensions of size $L$.
We first obtained a description which interpolated between the
expected behaviors of very large and very small black holes and
then showed that a (first order) phase transition, possibly signaled
by an outburst of energy, occurs in the system when the horizon shrinks
below $L$ from a larger value.
The phase transition is related to both a change in the topology of the
horizon and the breaking of translational symmetry along extra
dimensions.
In the present work we extend the results of \cite{ch} by investigating
several factors which affect the decay rate of a black hole.
\par
In Section~\ref{potential} we review and improve on the effective
potential discussed in Ref.~\cite{ch}.
In Section~\ref{evaporation} we then analyze the evaporation of
a black hole in $4+d$ dimensions both from the canonical and the
microcanonical point of view and show that the luminosity of small
black holes, such as primordial black holes (PBHs) which might be
sources of $\gamma$-ray bursts, strongly depends on the number of
space-time dimensions.
Such a dependence is further enhanced by the analysis of wave
propagation in Section~\ref{KG} where we argue that the heighth
of the potential barrier which affects the grey-body factor depends
on $d$.
The phase transition and the accompanying topology change which a black
hole undergoes as its horizon radius becomes of the order of the size of
the extra dimensions ($R_H \sim L$) are discussed in Section IV.
Finally, in Section~\ref{conc} we summarize and discuss our results.
We use units with $c=\hbar=1$.
\section{The Effective Potential}
\setcounter{equation}{0}
\label{potential}
If large extra spatial dimensions exist in nature, deviations from
Newton's law will be detected at the scale of the extra dimensions.
Assuming that all of the matter described by the standard model lives
on a four-dimensional D3-brane, the form of Newton's law can be obtained
for a point-like mass by means of Gauss' law \cite{arkani}.
Denoting by $r$ the radial distance in $4+d$ dimensions and by
$r_b$ the radial distance as measured on the D3-brane, we
find for distances $r$ much greater than the typical size of the extra
dimension $L$ a potential of the form
\be
V_{(4)}=-G_N\,{M\over r_b}
\ ,
\label{V>}
\ee
where $G_N=m_p^{-2}$ is Newton's constant in four dimensions.
On the other hand for $r\ll L$ the potential becomes
\be
V_{(4+d)}=-G_{(4+d)}\,{M\over r^{1+d}}
\ ,
\label{V<}
\ee
with $G_{(4+d)}=M_{(4+d)}^{-2-d}=L^d\,G_N$.
This implies that the huge Planck mass
$m_p^2=M_{(4+d)}^{2+d}\,L^d$ and, for sufficiently large
$L$ and $d$, the bulk mass scale $M_{(4+d)}$ (eventually
identified with the fundamental string scale) can be as small
as $1\,$TeV.
Since
\be
L\sim \left[{1\,{\rm TeV}/M_{(4+d)}}\right]^{1+{2\over d}}\,
10^{{31\over d}-16}\,{\rm mm}
\ ,
\label{tev}
\ee
requiring that Newton's law not be violated for distances larger
than $1\,$mm restricts $d\ge 2$ \cite{arkani,long}.
Further bounds are obtained by estimating the production of
KK-gravitons and support higher values of $d$ \cite{bounds}.
\par
Since our purpose is to study the decay of black holes, we would
like to be able to describe the metric tensor elements of a space
surrounding a black hole for all values of the horizon $R_H$ of
the black hole, including $R_H \sim L$.
Black holes with both very large horizon radii ($R_H\gg L$) and
very small radii ($R_H\ll L$) have been extensively investigated.
In the former case, the compact extra dimensions can be unwrapped
and the real singularity can be regarded as spread along a (black)
$d$-brane \cite{horowitz} of uniform density $M/L^d$, thus obtaining
the Schwarzschild metric on the orthogonal D3-brane, in agreement with
the weak field limit $V_{(4)}$, and an approximate ``cylindrical''
horizon topology $S^2\times \real^d$.
In the latter case a solution is known \cite{chamblin}, for one
infinite extra dimension \cite{RS}, which still has the form of a
black string extending all the way through the bulk AdS$_5$.
However, this solution is unstable \cite{gregory} and believed to
further collapse into one point-like singularity \cite{chamblin}.
This can be also argued from the observation that the Euclidean
action of a black hole is proportional to its horizon area and is
thus minimized by the spherical topology $S^{2+d}$.
Hence, small black holes are expected to correspond to a
generalization of the Schwarzschild metric to $4+d$ dimensions
\cite{myers} and should be colder and (possibly much) longer
lived \cite{argyres}.
\par
As a black hole evaporates the topology of the horizon changes from
$S^2 \times R^d$ to $S^{2+d}$ when $R_H$ decreases from $R_H>L$
to $R_H<L$.
For small black holes a complete description would provide an explicit
matching between the cylindrical metric (for $r\gg L$) and the
spherical metric (for $r\ll L$).
This is not a trivial detail, since the ADM mass of a spherical
$4+d$ dimensional black hole is zero as seen from the D3-brane
because there is no $1/r_b$ term in the large $r$ expansion of
the time-time component of the metric tensor \cite{myers}.
Thus, one concludes that the four-dimensional ADM mass $M$ of a
small black hole can be determined as a function of the horizon
radius $R_H$ (which governs the evaporation) only after such a
matching is provided explicitly.
One can further guess that $M$ emerges either as a
{\em boundary effect} ({\em i.e.},
due to the size of the extra-dimension being bounded) or as
a consequence of a non-zero tension (energy density) on the
D3-brane (this latter possibility will not be considered here).
Even less is known about black holes of size $R_H\sim L$, and a
complete description is likely to be achieved only by solving the
entire set of field equations for an evaporating black hole
in $4+d$ dimensions.
\par
\begin{figure}
\centering
\raisebox{6cm}{${r_b}$}\hspace{-0.0cm}
\epsfxsize=10cm
\epsfbox{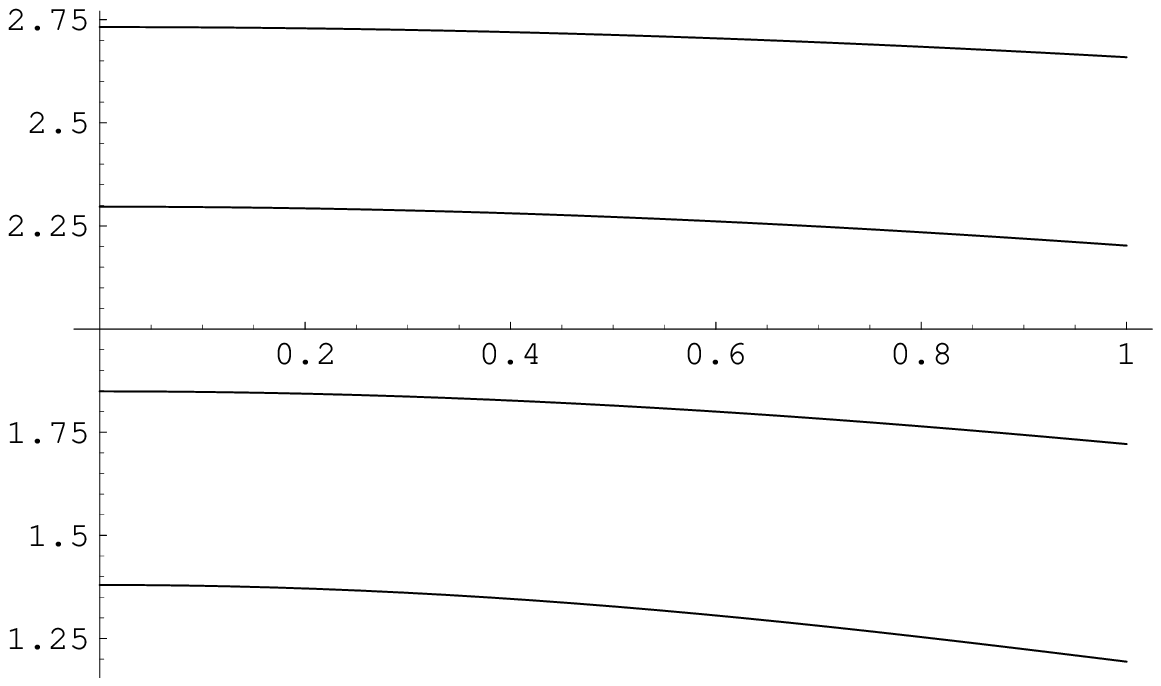}
\raisebox{3cm}{$y$}
\caption{Plot of the horizon for $d=1$ and $M=M_c$, $0.8\,M_c$,
$0.6\,M_c$, $0.4\,M_c$ (in units with $G=L=1$).
For $d=0$ the corresponding horizons would be at $r_b=2$, $1.6$, $1.2$, $0.8$.}
\label{RHd1}
\end{figure}

Rather than attempting the formidable problem of obtaining a
complete description of an evaporating black hole from the field
equations, we approximate the (time and radial components of the)
metric in $4+d$ dimensions as
\be
g_{tt}&\simeq& -1-2\,V(r)
\nonumber \\
&&\label{met}
\\
g_{rr}&\simeq&-g_{tt}^{-1}
\nonumber\ ,
\ee
where
\be
V(r)=-G_N\,{M\over r_b}\,\Theta(r_b-L)-
\sum_{n=1}^d\,
{C_n\,L^n\,G_N\,M\,\Theta(L-r_b)\over
\left[r_b^2+D_n\,\Theta(L-r_b)\,\sum_{i=1}^d\,y_i^2\right]^{(1+n)/2}}
\ ,
\label{pot}
\ee
where $\Theta$ is the Heaviside function (or a smooth approximation
of it), the $y_i$s are cartesian coordinates in the $d$ extra
dimensions and $C_n$, $D_n$ are numerical coefficients.
This yields $M$ as the ADM mass (on the D3-brane) of the black hole and
the radius of the horizon is determined by
\be
g_{rr}^{-1}=0
\ .
\label{R_H}
\ee
The above {\em ansatz\/} does not provide an exact solution of
the vacuum Einstein equations, since some of the components
of the corresponding Einstein tensor in $4+d$ dimensions
$G_{ij}=8\,\pi\,G_{(4+d)}\,T_{ij}\not =0$.
However, the coefficients $C_n$ and $D_n$ can be chosen in such
a way that the ``effective matter contribution'' $T_{ij}$ from
the region outside the black hole horizon is small.
In particular, one can require the contribution to the ADM
mass to be negligible,
\be
m&\equiv&\int_{R_H}^\infty d^{4+d}x\,T^t_{\ t}
\nonumber\\
&=&{1\over 8\,\pi\,G_{(4+d)}}\,
\int_{R_H}^\infty d^{4+d}x\,G^t_{\ t}(\{C_n\})
\ll M
\ .
\ee
In this sense one can render the above metric a good approximation
to a true black hole in $4+d$ dimensions.
For instance, for $d=1$, we obtain
\be
V(r)=\left\{\begin{array}{ll}
-G_N\,\strut\displaystyle{M\over r_b} & r>a\,L\gg L \\
& \\
-G_N\,\strut\displaystyle{M\over r_b}
-C\,\strut\displaystyle{L\,G_N\,M\over r_b^2+y^2}
\ \ \ \
& L<b\,L<r<a\,L \\
& \\
-C\,\strut\displaystyle{L\,G_N\,M\over r_b^2+y^2} &
r<b\,L
\ ,
\end{array}\right.
\ee
where the various coefficients must satisfy
\be
{m\over M}=C\,{a-b\over a\,b}\ll 1
\ ,
\ee
which determines the size of the region in which both $1/r$ and
$1/r^2$ terms are turned on \cite{smooth}.
Correspondingly one finds
\be
R_H(y=0)=\left\{\begin{array}{ll}
2\,G_N\,M & M\gg M_c \\
& \\
G_N\,M\,\left(1+\sqrt{1+\strut\displaystyle{2\,C\,L\over G_N\,M}}\right)
\ \ \ & M\sim M_c \\
& \\
\sqrt{2\,L\,G_N\,M} & M\ll M_c
\ ,
\end{array}\right.
\ee
where $M_c$ is the value of $M$ for which the usual four-dimensional
horizon radius $2\,G_N\,M=L$ [the value of $M$ for which $R_H=L$
can be obtained only after solving Eq.~(\ref{R_H}), which is an algebraic
equation of order $d+2$].
For $M\sim M_c$ the radius of the horizon also depends on the coordinate
$y$ and is neither cylindrically nor spherically symmetric (see Fig.~\ref{RHd1}).
Analogous estimates can be obtained for $d>1$.
\par
Finally, we note that, although the ansatz in Eq.~(\ref{met})
is not a true vacuum solution, such a solution would not be very
useful for $R_H\ll L$ in any event, because once Hawking radiation
is included, its backreaction on the metric at small $r$ is likely
to be significant.
Therefore, it is sufficient to impose the condition that $m$ be comparable
with the energy density of Hawking's radiation integrated from
$r=R_H$ to $r=\infty$, {\em i.e.} the amount of mass lost into Hawking
particles.
This quantity of course depends on the initial mass of the black hole.
\section{Black Hole Evaporation}
\setcounter{equation}{0}
\label{evaporation}
There are several factors affecting the rate of black hole
evaporation due to Hawking radiation.
We begin with a review of the expression for the decay as obtained
from the canonical ensemble and then discuss the modifications to
this expression which arise when the more basic microcanonical
ensemble is used.
Finally, we estimate the grey-body factor by studying the
Klein-Gordan equation.
\subsection{The canonical picture}
When the size of the black hole is large compared to the extra
dimensions ($R_H \gg L$), $r \simeq r_b$ and the metric
(\ref{met}) is approximately cylindrically symmetric (along the
extra dimensions).
Further, when $L$ is set to $0$, $m=0$, and Eq.~(\ref{R_H})
coincides with the usual four dimensional Schwarzschild radius
\be
R_H\simeq 2\,\ell_p\,{M\over m_p}
\ .
\label{R_H>}
\ee
Correspondingly the inverse Hawking temperature
($\beta_H=T_H^{-1}$) and Euclidean action are \cite{hawking,st}
\be
\beta_H^>&\simeq& 8\,\pi\,\ell_p\,{M\over m_p}
\label{B>}
\\
\nonumber \\
S_E^>&\simeq&4\,\pi\,\left({M\over m_p}\right)^2
\simeq{{\cal A}_{(4)}\over 4\,\ell_p^2}
\ ,
\label{SE>}
\ee
where ${\cal A}_{(D)}$ is the area of the horizon in $D$ space-time
dimensions and the condition $R_H\gg L$ translates into
\be
M\gg m_p\,{L\over\ell_p}\equiv M_c
\sim \left({L\over 1\,{\rm mm}}\right)\,10^{27}\,{\rm g}
\ .
\label{Mc}
\ee
The fact that the extra dimensions do not play any significant role at
this stage is further confirmed by $T_H^>\ll 2\,\pi/L$.
Since $2\, \pi/L \equiv m^{(1)}$ is the mass of the lightest
Kaluza-Klein(KK) mode, no KK particles can be produced.
\par
The energy loss per unit time for an evaporating black hole is given by
\be
{dM\over dt}=-{F}_{(4+d)}^>\simeq-{F}_{(4)}^>
\ ,
\ee
where ${F}_{(D)}$ is the total luminosity as measured in $D$
space-time dimensions.
Since microcanonical corrections have been shown to become significant
only for $M\sim m_p$ (in four space-time dimensions \cite{mfd}), one can
further approximate the luminosity by employing the (simpler) canonical
expression \cite{hawking}
\be
{F}_{(D)}&\simeq&{\cal A}_{(D)}\,
\sum_s\,\int_0^\infty {\Gamma_s(\omega)\,\omega^{D-1}
\over e^{\beta_H\,\omega}\mp 1}\,d\omega
\nonumber \\
&=&{\cal A}_{(D)}\,N_{(D)}\,(T_H)^D
\ ,
\label{L}
\ee
where $\Gamma$ is the grey-body factor and $N$ is a coefficient which
depends upon the number of available particle species $s$ with energy
smaller than $T_H$ and is also affected by the value of $\Gamma$.
For $D=4$ and $\beta_H=\beta_H^>$ one obtains the well known result
\cite{hawking}
\be
{F}_{(4)}^>\sim {N_{(4)}\over \ell_p
^2}\,\left({m_p\over M}\right)^2
\ .
\label{L4>}
\ee
\par
For $R_H \ll L$ the parameters in the potential (\ref{pot})
can be chosen to minimize the contribution of $m$ to the ADM mass
and this is generally obtained by switching off all terms except
that going as $1/r^{1+d}$ for $r<L$.
Eq.~(\ref{R_H}) then leads to
\be
R_H\simeq \left(2\,L^d\,G_N\,M\right)^{1/1+d}
\ ,
\label{R_H<}
\ee
and the consistency conditions $\ell_p\ll R_H\ll L$ hold for
\be
m_p\,\left({\ell_p\over L}\right)^{d\over 2+d}\ll M\ll M_c
\ .
\ee
Since we have assumed that the spherical symmetry extends to
$4+d$ dimensions, one obtains \cite{argyres}
\be
&&\beta_H^<\sim L\,\left({M\over M_c}\right)^{1\over 1+d}
\\
\nonumber \\
&&S_E^<\sim\left({L\over\ell_p}\right)^2\,
\left({M\over M_c}\right)^{2+d\over 1+d}
\sim{A_{(4+d)}\over\ell_p^2\,L^d}
\ ,
\label{SE<}
\ee
which reduce back to (\ref{B>}) and (\ref{SE>}) if one pushes
down $L\to \ell_p$ ($M_c\to m_p$).
For $M\ll M_c$, the temperature $T_H^<$ is sufficient to excite KK
modes, although it is lower than that of a four-dimensional black
hole of equal mass.
Correspondingly, the Euclidean action $S_E^<(M)\ge S_E^>(M)$,
yielding a smaller tunnelling probability \cite{hawking,r1}
\be
P\sim\exp\left(-S_E\right)
\ ,
\label{P}
\ee
that is, a smaller probability for the Hawking particles to be created
in the $4+d$ dimensional scenario.
\par
Although ordinary matter is confined on the D3-brane, a black hole can
emit particles via Hawking's process into all of the $3+d$ spatial
directions of the bulk.
For $R_H \ll L$ the evaporation of a black hole can be obtained
from Eq.~(\ref{L}) with $D=4+d$ and $T_H=T_H^<$,
\be
{dM\over dt}=-{F}_{(4+d)}^<
\sim
{N_{(4+d)}\over L^2}\,\left({M_c\over M}\right)^{2\over 1+d}
\ .
\label{L<}
\ee
The luminosity (\ref{L}) can be written as
\be
F_{(D)}={N_{(D)}\over R_H^2}
\ .
\ee
Thus, a comparison of the radius of the (apparent) horizon (\ref{R_H<})
with the analogous quantity in Eq.~(\ref{R_H>}) shows that
the luminosity of a black hole of given ADM mass $M<M_c$ is much
smaller in $4+d$ dimensions than it would be with no extra dimensions.
This assertion might have to be qualified if the $N_{(D)}$ term in
the luminosity were greater in $4+d$ dimensions than the one in four
dimensions.
We show below that this is not the case.
\par
Conservation of energy in $4+d$ dimensions requires that
\be
F_{(4+d)} = F_{(4)} + F_{KK}
\ee
and since $F_{(D)}\sim N_{(D)}$ [Eq.~(\ref{L})] and the number of
degrees of freedom of KK gravitons is much less than the number of
standard model particles ($N_{KK}\ll N_{(4)}$), the energy emitted
into the KK modes must be a small fraction of the total luminosity
(similar results were obtained in Ref.~\cite{emparan})
\be
{F_{KK}\over F_{(4+d)}}\sim
{N_{KK}\over N_{(4)}+N_{KK}}\ll 1
\ .
\ee
Standard model particles on the D3-brane with high enough energy
({\em e.g.}, larger than $\Lambda_{EW}$) might be able to overcome
the confining mechanism.
In this case the bulk standard model fields should be included
among the KK modes and $N_{KK}$ should be considered as an
increasing function of the temperature.
In this scenario the ratio $F_{KK}/F_{(4+d)}$ would eventually
reach unity when $T_H^< \gg\Lambda_{EW}$.
\subsection{The Microcanonical Picture}
\label{Smicro}
The canonical ensemble description used above is in principle
incorrect because a black hole in asymptotically flat space-time
cannot be in thermal equilibrium with its Hawking radiation.
However for large black holes this description is a very good
approximation to the true picture.
The correct statistical mechanical description of black holes
utilizes the microcanonical ensemble \cite{r1,mfd}.
\par
\begin{figure}
\centering
\raisebox{6cm}{${dM\over dt}$}\hspace{-0.0cm}
\epsfxsize=10cm
\epsfbox{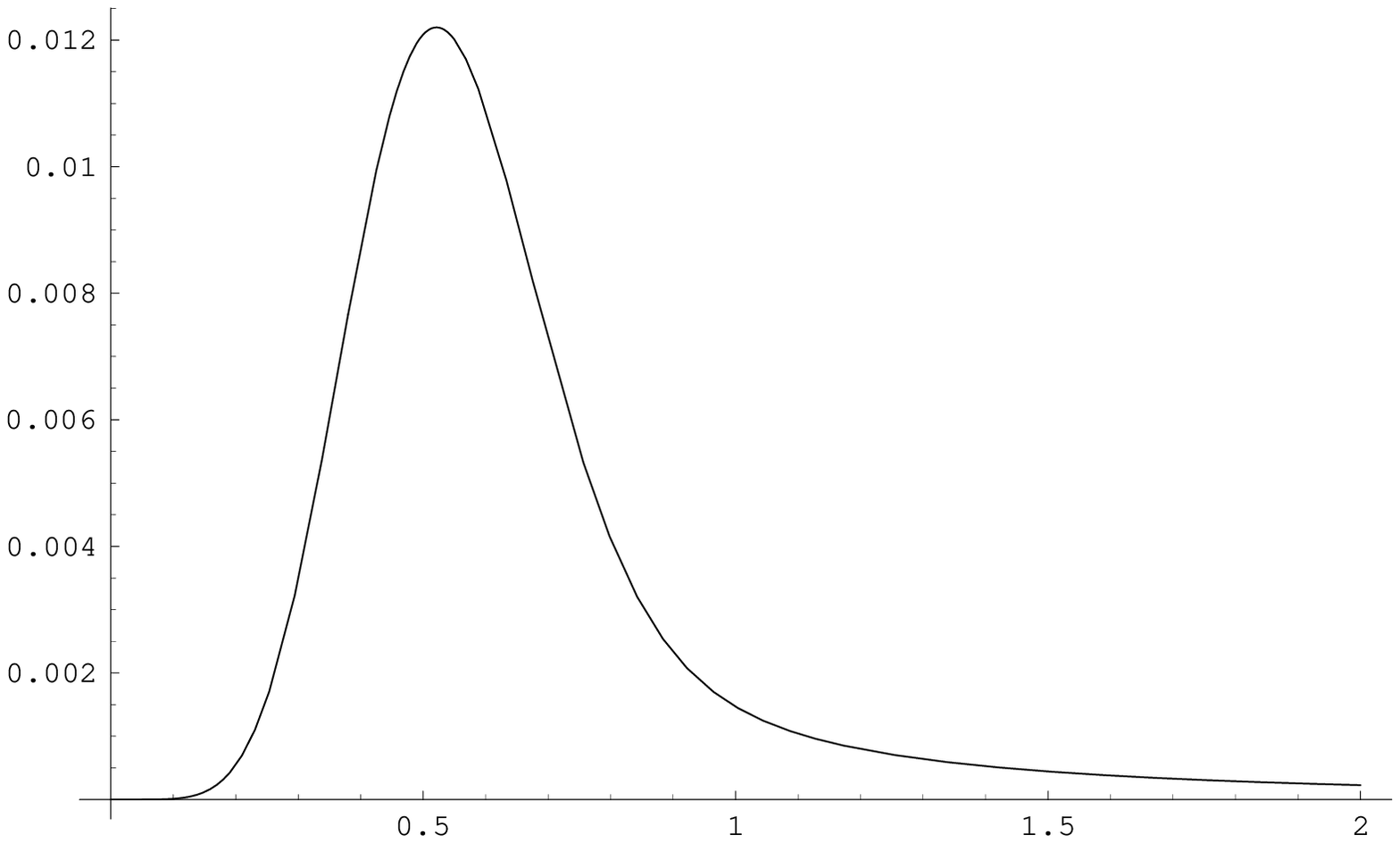}\\
{\hspace{6cm} $M/m_p$}
\caption{Decay rate for a large ($R_H\gg L$) black hole.}
\label{microdMdt_l}
\end{figure}

For $R_H \gg L$ the topology of the horizon is ``cylindrical'',
and the Euclidean action is
\be
S^>_E \simeq 4\,\pi\,\left({M\over{m_p}}\right)^2
\ .
\ee
The number density in the microcanonical enesemble for this case is
\be
n^>(\omega)=\sum_{l=1}^{[[M/\omega]]}\,
{\exp\left[4\,\pi\,(M-l\,\omega)^2/m_p^2\right]
\over{\exp(4\,\pi\,M^2/m_p^2)}}
\ ,
\ee
where $[[X]]$ denotes the integer part of $X$.
In the limit $M \to \infty$, $n^>$ is equal to the canonical ensemble
number density used in Eq.~(\ref{L}).
The decay rate for a black hole in $4+d$ dimensions is given by
\be
{dM\over dt}={\cal A}_{(4+d)}\,\int_0^{\infty}
n\,(\omega)\,\Gamma(\omega)\,\omega^{d+3}\,d\omega
\ .
\label{dMdt}
\ee
Making the substitutions $n=n^>$, $x=M-l\,\omega$ and assuming that
$\Gamma\simeq 1$, the decay rate in four dimensions can be
written as
\be
{dM\over dt}={\cal A}_{(4)}\,I\,\sum_{l=1}^{\infty}{1\over{l^4}}
={\pi^4\over{90}}\,{\cal A}_{(4)}\,I
\ ,
\ee
where
\be
I=\int_0^M {(M-x)^3\,\exp\left(4\,\pi\,x^2/m_p^2\right)\,dx
\over{\exp\left(4\,\pi\,M^2/m_p^2\right)}}
\ .
\ee
A plot of the decay rate as a function of the mass is shown in
Fig.~\ref{microdMdt_l}.
However, we remark that, in the presence of extra dimensions,
this description holds only for $M>M_c$ ($R_H>L$), therefore
the plot in Fig.~\ref{microdMdt_l} is probably not accurate for
small masses.
It should in fact be considered only in the region $M/m_p>L/\ell_p$
where there is no essential difference between the microcanonical
and the canonical luminosities (\ref{L4>}).
\par
\begin{figure}
\centering
\raisebox{6cm}{${dM\over dt}$}\hspace{-0.0cm}
\epsfxsize=10cm
\epsfbox{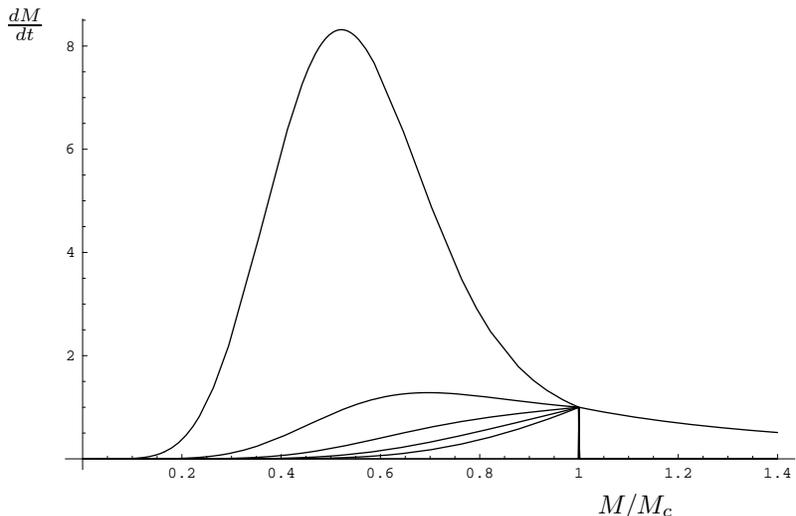}\\
{\hspace{6cm} $M/M_c$}
\caption{Decay rate for a small black hole in increasing number of extra
dimensions ($d=0$ uppermost curve, $d=4$ lowest curve).
Vertical units are arbitrary.}
\label{microdMdt_s}
\end{figure}
\begin{figure}
\centering
\raisebox{6cm}{${M\over M_c}$}\hspace{-0.0cm}
\epsfxsize=10cm
\epsfbox{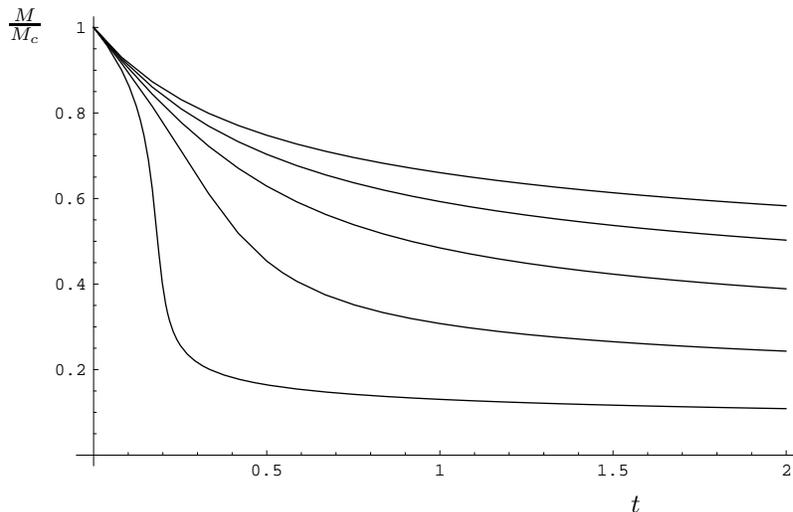}\\
{\hspace{6cm} $t$}
\caption{History of a small [$M(0)\sim M_c$] black hole in increasing number
of extra dimensions ($d=0$ lowest curve, $d=4$ uppermost curve).
Horizontal units are arbitrary.}
\label{microdMdt_st}
\end{figure}

For $R_H<L$ the Euclidean action is
\be
S_E^< =4\,\pi\,\left({L\over{\ell_p}}\right)^2\,
\left({M\over{M_c}}\right)^{(d+2)/(d+1)}
\ ,
\ee
where $M_c=m_p\,L/\ell_p$.
The number density for horizon radii of this size is
\be
n^<(\omega)=\sum_{l=1}^{[[M/\omega]]}\,
{\exp\left[4\,\pi\,\left({M-l\,\omega\over{M_c}}\right)^{d+2\over d+1}
\,\left({L\over{\ell_p}}\right)^2\right]
\over{\exp\left[4\,\pi\,\left({M\over{M_c}}\right)^{d+2\over d+1}\,
\left({L\over{\ell_p}}\right)^2\right]}}
\ .
\ee
The decay rate corresponding to this number density as calculated
from Eq.~(\ref{dMdt}) is exhibited for $d=0,\ldots,4$ in
Fig.~\ref{microdMdt_s}.
In this plot, we have matched the microcanonical decay rates for $M<M_c$
with the canonical rate given in Eq.~(\ref{L4>}) for $M>M_c$ by
multiplying the former by a suitable constant in order to show that
there is a peak in the luminosity around $M_c$.
In fact, the curves show clearly that the decay rates for all cases with
$d>1$ have a maximum at $M\simeq M_c$ and black holes with masses less
than $M_c$ thus have a significantly slower evaporation rate which results
in a longer lifetime (see Fig.~\ref{microdMdt_st}).
The case $d=1$ appears slightly different, but
this value of $d$ must be rejected because it leads to contradictions
with experimental tests of Newton's law \cite{arkani,long}.
We wish to make two remarks about the above estimates:
firstly, we expect the precise shapes of the decay rates in
Fig.~\ref{microdMdt_s} at $M\sim M_c$ depend on the details of the phase
transition which we are able to describe only qualitatively
(see Section~\ref{topology}).
As a consequence, the curves in Fig.~\ref{microdMdt_st} are not totally
reliable around $t=0$ ($M\sim M_c$), but can be used to describe black
holes which are produced with $M(0)$ significantly less than $M_c$;
secondly, we have always set $\Gamma\sim 1$, but the grey-body factor
actually depends strongly on the number of extra dimensions $d$ causing
a further suppression of the decay rate, as we show in Section~\ref{KG}.
\par
A reduced decay rate for black holes in large extra dimensions has
several interesting implications for cosmological phenomena involving
PBHs.
For $d=2$ and $M_{(6)}\sim 1\,$TeV, the size of the extra dimension
is $L\sim 1\,$mm [from Eq.~(\ref{tev})] and $M_c\sim10^{27}\,$g
[from Eq.~(\ref{Mc})].
Since the initial mass of an astrophysical black hole is known to
exceed at least a few solar masses ($M_0>10^7\,M_c$) \cite{astrobh},
its $T_H^>$ is cooler than the cosmic background radiation and it
cannot decay to $M_c$ in any reasonable amount of time.
However, PBHs (see, {\em e.g.}, Refs.~\cite{pbh}) could have been
produced with an initial mass $M_0\ll M_c$, so that $R_H\ll L$ from
the very beginning and, according to Fig.~(\ref{microdMdt_s}) the
lifetimes of such black holes would be many orders of magnitude
longer than those living in four spacetime dimensions.
For instance, a PBH with $M_0\sim 10^{15}\,$g, which has a life-time
comparable with the age of the Universe according to Eq.~(\ref{L4>})
(the canonical ensemble is a good approximation in this mass range
for $d=0$),
in this new scenario would decay in $\tau_{(6)}\sim 10^{16}\,$ times
the age of the Universe.
On the other hand, if the PBH had been produced with $R_H>L$, we can
speculate that, after it shrank down to near $L$ according to
Eq.~(\ref{L4>}), either the evaporation continued and its mass
eventually entered into the regime of very slow decay, or the
evaporation stopped (because of the backreaction) at $M\sim M_c$.
The latter option is even more dramatic, since it would leave remnants
of mass as large as $M_c$ (to be compared with possible remnants of
Planck size \cite{pbh} as follows from microcanonical estimates in
four dimensions \cite{mfd}).
\par
One can therefore highlight a list of topics involving PBHs upon which
models with extra dimensions might have a bearing \cite{argyres}:
\begin{enumerate}
\item
PBHs could be the source of $\gamma$-ray bursts and cosmic rays only if
their initial mass was such that $M\ll 10^5\,$g \cite{age} or $M\sim M_c$
at the time of the emission;
the allowed density of PBHs which fits into bounds obtained from current
observations could then be significantly changed, with consequences on
\item
models of the early Universe,
\item
dark matter (in the form of KK particles \cite{arkani}) and
\item
density and size of microlensing sources \cite{microlensing}.
\end{enumerate}
\subsection{Angular Momentum Barrier to Decay}
\label{KG}
A scalar wave in $4+d$ dimensions satisfies the equation
\be
\Box\,\Phi=0
\ ,
\label{kg}
\ee
where the D'Alembertian is given by
\be
\Box={1\over\sqrt{-g}}\,\partial_\mu\,\left(
\sqrt{-g}\,g^{\mu\nu}\,\partial_\nu\right)
\ ,
\ee
with (in spherical coordinates for which $\theta_1 = \theta$ and
$\theta_2=\phi$)
\be
\sqrt{-g}=r^{2+d}\,\prod_{i=1}^{d+2}(\sin\theta_i)^{d+2-i}
\ee
\par
For $L\ll R_H<r$ we can neglect the extra dimensions and simply take
the standard Schwarzschild line element on the brane,
\be
ds^2\simeq-\Delta_{(4)}\,dt^2+{dr^2\over \Delta_{(4)}}+
r^2\left(d\theta^2+\sin^2\theta\,d\phi^2\right)
\ ,
\ee
where
\be
\Delta_{(D)}=1-\left({R_H\over r}\right)^{D-3}
\ .
\ee
\par
For $R_H<r\ll L$ one can analogously consider a spherically
symmetric black hole in $4+d$ dimensions \cite{myers}.
However, in order to take into account the fact that $d$ spatial
dimensions have size $L$, we shall instead use the following form
\be
ds^2\simeq-\Delta_{(4+d)}\,dt^2+{dr^2\over \Delta_{(4+d)}}+
r^2\,\left(d\theta^2+\sin^2\theta\,d\phi^2\right)
+r^2\,\sum_{i=1}^d\,d\phi_i^2
\ ,
\label{g4d}
\ee
where $R_H<r<L$ is the $4+d$ dimensional areal coordinate and
$dx^i\simeq r\,d\phi_i$ are cartesian coordinates in the
extra dimensions.
\par
We assume the scalar field $\Phi$ can be factorized according to
\be
\Phi=e^{i\,\omega\,t}\,
R(r)\,S(\theta)\,e^{i\,m\,\phi}\,e^{i\,\sum\,n_i\,\phi_i}
\ ,
\ee
with $n_i$ positive integers, so that $\Phi$ satisfies periodic
boundary conditions at the edges of the bulk ($y_i=\pm L/2$).
The radial equation obtained from Eq.~(\ref{kg}) for the metric
(\ref{g4d}) then becomes ($\Delta\equiv\Delta_{(4+d)}$)
\be
\left[-{\Delta\over r^{2+d}}\,{d\over dr}\,
\left(r^{2+d}\,\Delta\,{d\over dr}\right)
+\omega^2
-{\Delta\over r^2}\,A\right]\,R=0
\ee
and the angular equation is
\be
\left[{1\over\sin\theta}\,{\partial\over\partial\theta}\,\left(
\sin\theta\,{\partial\over\partial\theta}\right)
-{m^2\over\sin^2\theta}\right]\,S=\lambda\,S
\ ,
\ee
where $A=\lambda+\sum\,n_i^2$, the separation constant
$\lambda=l\,(l+1)$ and $S=Y_l^m$ is a standard spherical
harmonic.
\par
The radial equation can be further simplified by defining a
tortoise coordinate
\be
dr_*\equiv{dr\over\Delta}
\ ,
\ee
and introducing a rescaled radial function,
\be
W\equiv r^{1+d/2}\,R
\ ,
\ee
which then satisfies a Schr\"odinger-like equation
\be
\left[-{d^2\over dr_*^2}+V\right]\,W=\omega^2\,W
\ ,
\ee
where the potential $V=V_1+V_2$ is given by
\be
&&V_1=\left[(1+d)\,\left(1+{d\over 2}\right)+ A\right]\,
{\Delta\over{r^{2}}}
\nonumber \\
\\
&&V_2=-\,\left(1+{d\over 2}\right)^2\,{\Delta^2\over r^2}
\ .
\nonumber
\ee
\par
\begin{figure}
\centering
\raisebox{6cm}{$V$}\hspace{-0.0cm}
\epsfxsize=10cm
\epsfbox{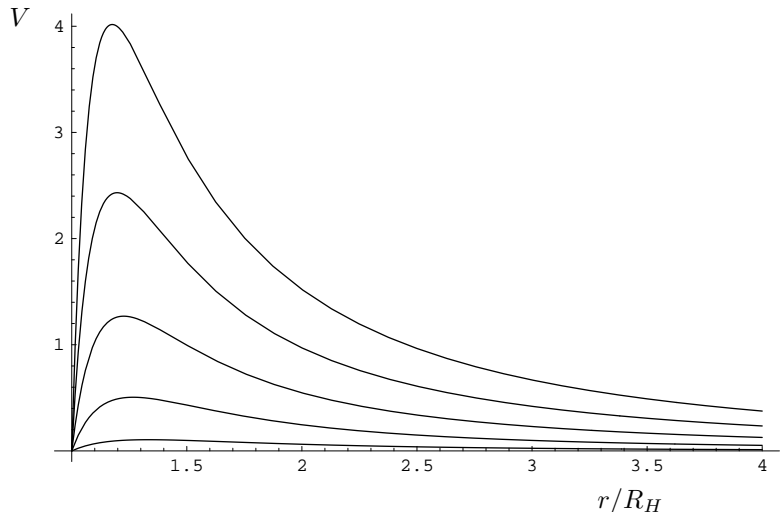}\\
{\hspace{6cm} ${r/R_H}$}
\caption{Potential $V_1+V_2$ for different numbers of extra dimensions
($d=4$ uppermost curve, $d=0$ lowest curve).}
\label{KGV}
\end{figure}

The contribution $V_2$ vanishes sufficiently fast near $R_H$ and
is negligible there.
The potential $V_1$ generates a barrier located outside
the horizon which suppresses the grey-body factor for all modes of
the scalar field including those with zero angular momentum
\cite{birrell}.
The effect for $d=0$ is mild, however for $d>0$ the barrier increases,
significantly reducing the black hole decay rate.
From the plot of the potential in Fig.~\ref{KGV} one can estimate the
(frequency dependent) suppression factors with respect to the purely
four-dimensional case (for which $\Gamma\sim 1$) by making use of the
W.K.B. approximation for the transmission probability
\be
\Gamma(\omega)\sim\exp\left(-2\,\int dr\,\sqrt{\left|V-\omega\right|}\right)
\ ,
\ee
where the integral must be performed between the two values of $r$
at which $V=\omega$ (for $\omega$ smaller than the maximum of $V$).
For the typical frequency $\omega\sim T_H^<$ ($\sim 1$ in the units of
Fig.~\ref{KGV}) one obtains
\be
&&\Gamma_{d=1}\sim 1
\nonumber \\
&&\Gamma_{d=2}\sim 0.73
\nonumber \\
&&\Gamma_{d=3}\sim 0.22
\nonumber \\
&&\Gamma_{d=4}\sim 0.05
\ .
\label{gammas}
\ee
One then observes that, for frequency $\omega<T_H^<$ the transmission
probability is smaller, while for quanta of frequency $\omega\gg T_H^<$
the probability of creation is actually negligible.
We remark that the curves shown in Fig.~\ref{KGV} were obtained by
setting $A=0$ ({\em i.e.}, $l=n_i=0$) and, since the $n_i$s cannot be
zero, the curves represent lower bounds on the heighths of the
curves and the values in Eq.~(\ref{gammas}) represent upper bounds on the
grey-body factors.
This barrier effect taken together with the decay rate reduction
for $d>1$ (again, we recall that the case $d=1$ is ruled out) as shown
in Fig.~\ref{microdMdt_s} would seem to suggest that black holes with
horizon radii on the order of the size of the extra dimensions or less
decay very little if at all.
\section{Horizon topology change and phase transition}
\begin{figure}
\centering
\epsfxsize=10cm
\epsfbox{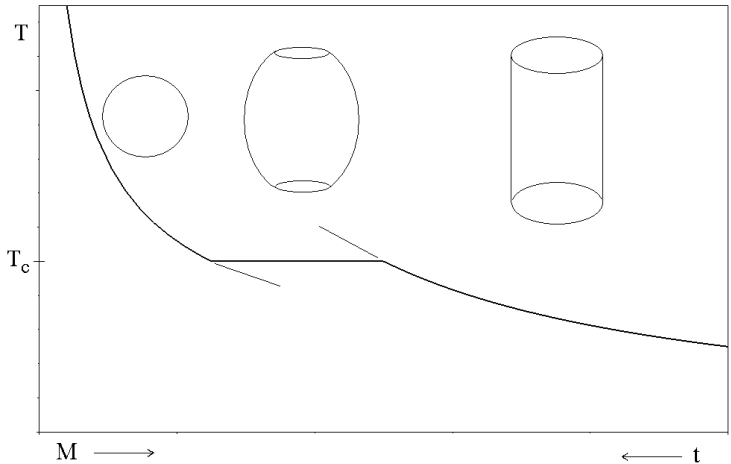}
\caption{Sketch of the phase transition described in the text.
Note that the shape of the horizon along the extra dimension(s) for
$T\sim T_c$ is shown in Fig.~\ref{RHd1} (for $d=1$).}
\label{fig2}
\end{figure}

\setcounter{equation}{0}
\label{topology}
In the foregoing analysis horizon radii on the order of the size of
the extra dimensions ($R_H \sim L$) have not been considered in great
detail.
For example, the expressions for $\beta_H$ and $S_E$ in this region
are discontinuous because they contain numerical coefficients which
depend on the topology and number of the extra dimensions.
For instance,
\be
S_E^<(M_c)\simeq {S_{(4+d)}\over S_{(4)}}\,S_E^>(M_c)
\ .
\label{ent}
\ee
A complete description of this process likely requires solving the full
quantum backreaction problem in the bulk when $M\sim M_c$, but we can get
further physical insight by appealing to the point of view of the
statistical mechanics.
Although the use of the canonical ensemble is incorrect, it still leads
to results in fairly good agreement with the microcanonical description
for black holes of mass greater than $m_p$ in four dimensions and
greater than $M_c$ in $4+d$ dimensions.
We can then introduce an effective partition function for the black hole
and its Hawking radiation as the Laplace transform of the microcanonical
density of states but with a cut-off $\Lambda$ which makes the partition
function integral finite,
\be
Z_\beta\sim\int^\Lambda dM\,e^{-\beta\,M}\,e^{S_E(M)}
\ .
\ee
It then follows from Eq.~(\ref{ent}) that there is a discontinuity in the
first derivative of the free energy $\beta\,F\equiv-\ln Z$ at
\be
\beta_c\sim L\sim {1\over m^{(1)}}
\ ,
\ee
that is a first order phase transition (see Fig.~\ref{fig2}):
In the cold phase, $T_H<\beta_c^{-1}$, the system appears ``condensed''
into the lower four dimensional 3-brane and translational invariance in
the $d$ extra directions is broken.
For $T_H>\beta_c^{-1}$ translational invariance is restored, and the
system starts spreading over all bulk space, with brane vibrations
playing the role of Nambu-Goldstone bosons which give mass to the KK
modes \cite{arkani}.
In this approximation the specific heats in the cold and hot phases are
given by
\be
&&C_V^>\sim -\left({M\over m_p}\right)^2
\\
&&C_V^<\sim -L\,M\,\left({M\over M_c}\right)^{1\over 1+d}
\ ,
\ee
Since $C_V^<$ is negative (as is $C_V^>$), one is eventually forced to
use the microcanonical description when the temperature is too high
\cite{mfd}, as we have done in Section~\ref{Smicro}.
\par
We end this Section by noting that the above analysis does not
determine the typical times of such a transition.
Using the results of the previous Section, one could
argue that the transition takes a very long time to occur
(if at all), since the luminosity of black holes is strongly damped
for $M<M_c$.
Although an outburst of energy at $M\sim M_c$ cannot be excluded
(and is actually favored according to the microcanonical rate
as displayed in Fig.~\ref{microdMdt_s}), it seems implausible that
the black hole can then complete the transition and reach very
small masses in a reasonable amount of time.
\section{Conclusions}
\setcounter{equation}{0}
\label{conc}
While the effective potential of Eq.~(\ref{met}) is not a vacuum
solution, it does have the desirable property of providing only a
small contribution to the ADM mass if the adjustable constants in
the potential are properly chosen.
It also contains a parameter which is readily identifiable as the
mass of the black hole for both $R_H>L$ and $R_H<L$.
\par
The analysis of the black hole decay rate in both the canonical and
microcanonical pictures shows that the presence of large extra
dimensions would slow the decay rate, and that the decay rate
decreases with increasing number of extra dimensions.
The source of this decrease in decay rate is the dependence of the
number density and the horizon area on the number of extra dimensions.
In addition to these effects the size of the angular momentum barrier
increases with the number of extra dimensions, further decreasing the
decay rate.
This result suggests that, even without taking backreaction effects
into account, black holes with horizon radii less than the size of the
extra dimensions are quasi-stable.
Thus, even small primordial black holes (masses less than
$M_0\sim 10^{15}\,$gm) could have lifetimes comparable to or even
much longer than the age of the present universe.
The detection of PBHs with masses less than $M_0$ would be evidence
for the existence of large extra dimensions, and the values of the
PBH masses would set limits on the number of extra dimensions.
\par
A statistical mechanical analysis of a black hole whose horizon
radius $R_H$ is approximately equal to the size $L$ of the extra
dimensions shows that as $R_H$ shrinks below $L$, a phase transition
occurs.
This event is accompanied by a change in the horizon topology from
``cylindrical'' ($S^2\times R$) to ``spherical'' ($S^3$).
An interesting extension of the present work would be to see if the
phase transition can be described in terms of the topology change.
\acknowledgments
This work was supported in part by the U.S. Department of Energy under
Grant no.~DE-FG02-96ER40967 and by NATO grant no.~CRG.973052.

\begin{thebibliography}{99}
%
%
%
\bibitem{arkani}
N. Arkani-Hamed, S. Dimopoulos and G. Dvali, Phys. Lett. {\bf B 429},
263 (1998); Phys. Rev. D {\bf 59}, 0806004 (1999);
I. Antoniadis, N. Arkani-Hamed, S. Dimopoulos and G. Dvali,
Phys. Lett. {\bf B 436}, 257 (1998).
%
\bibitem{RS}
L. Randall and R. Sundrum, Phys. Rev. Lett. {\bf 83}, 4690 (1999).
%
%
\bibitem{r1}
B. Harms and Y. Leblanc, Phys. Rev. D {\bf 46}, 2334 (1992);
Phys. Rev. D {\bf 47}, 2438 (1993);
Ann. Phys. {\bf 244}, 262 (1995);
Ann. Phys. {\bf 244}, 272 (1995);
Europhys. Lett. {\bf 27}, 557 (1994);
Ann. Phys. {\bf 242}, 265 (1995);
P.H. Cox, B. Harms and Y. Leblanc, Europhys. Lett. {\bf 26},
321 (1994).
%
\bibitem{mfd}
R. Casadio, B. Harms and Y. Leblanc, Phys. Rev. D {\bf 57}
1309 (1998);
R. Casadio and B. Harms, Phys. Rev. D {\bf 58}, 044014 (1998);
Mod. Phys. Lett. {\bf A17}, 1089 (1999).
J.M. Bardeen, B. Carter, S.W. Hawking, Commun. Math. Phys. {\bf 31},
161 (1973);
G.W. Gibbons and S.W. Hawking, Phys. Rev. D {\bf 15}, 2752 (1977).
%
\bibitem{polchinski}
J. Polchinski, TASI Lectures on D-branes, hep-th\-/\-97\-02\-136.
%
\bibitem{vafa}
A. Strominger and C. Vafa, Phys. Lett. B {\bf 379}, 99 (1996).
%
\bibitem{maldacena}
J. Maldacena, Nucl. Phys. Proc. Suppl. {\bf 61 A}, 111 (1998).
%
\bibitem{st}
J.D. Bekenstein, Phys. Rev. D {\bf 7}, 2333 (1973);
Phys. Rev. D {\bf 9}, 3292 (1974).
%
\bibitem{argyres}
P.C. Argyres, S. Dimopoulos and J. March-Russell, Phys. Lett.
{\bf B 441}, 96 (1998).
%
\bibitem{chamblin}
A. Chamblin, S. Hawking and H.S. Reall, Phys. Rev. D {\bf 61},
0605007 (2000).
%
\bibitem{emparan}
R. Emparan, G.T. Horowitz and R.C. Myers, hep-th/0003118.
%
\bibitem{ch}
R. Casadio and B. Harms, Phys. Lett. {\bf B 487}, 209 (2000).
%
\bibitem{long}
J.C. Long, H.W. Chan and J.C. Price, Nucl. Phys. {\bf B539}, 23
(1999).
%
\bibitem{bounds}
S. Cullen, M. Perelstein, Phys. Rev. Lett. {\bf 83}, 268 (1999);
V. Barger, T. Han, C. Kao and R.J. Zhang, Phys. Lett, {\bf B 461},
34 (1999);
M. Fairbairn, {\em Cosmological constraints on large extra dimensions},
hep-ph/0101131
%
\bibitem{horowitz}
G.T. Horowitz and A. Strominger, Nucl. Phys. {\bf B 360}, 197
(1991).
%
\bibitem{gregory}
R. Gregory and R. Laflamme, Phys. Rev. Lett. {\bf 70}, 2837 (1993).
%
\bibitem{myers}
R.C. Myers and M.J. Perry, Ann. Phys. {\bf 172}, 304 (1986).
%
\bibitem{smooth}
There are further contributions to $m$ from the ``strips'' $r\sim b\,L$ and
$r\sim a\,L$, which can be controlled by keeping $a\not=b$ and appropriately
``smoothing'' the $\Theta$s.
%
\bibitem{hawking}
S.W. Hawking, Nature {\bf 248}, 30 (1974);
Comm. Math. Phys. {\bf 43}, 199 (1975).
%
\bibitem{page}
D. Page, Phys. Rev. D {\bf 13}, 198 (1976);
Phys. Rev. D {\bf 16}, 2401 (1977).
%
\bibitem{alex}
S.L. Dubovsky, V.A. Rubakov and P.G. Tinyakov, hep-th/0006046;
R. Casadio, A. Gruppuso and G. Venturi, Phys. Lett. {\bf B 495},
378 (2000).
%
\bibitem{huang}
K. Huang, {\em Statistical mechanics}
(John Wiley and Sons, New York, 1987).
\bibitem{astrobh}
A. Celotti, S.C. Miller and D.W. Sciama, Class. Quantum Grav. {\bf 16}, A3 (1999).
%
\bibitem{pbh}
K. Kohri and J. Yokoyama, Phys. Rev. D {\bf 61}, 023501 (2000);
B.J. Carr and J.H. MacGibbon, Phys. Rep. {\bf 307}, 141 (1998);
F. Halzen, E. Zas, J.H. MacGibbon and T.C. Weeks, Nature {\bf 353}, 807
(1991).
%
\bibitem{age}
This is the initial mass of a PBH with $\tau_{(6)}$ of the order of the age
of the Universe.
%
\bibitem{microlensing}
C. Alcock {\em et al.}, {\em The MACHO project: microlensing results from
5.7 years of LMC observations}, astro-ph/0001272;
E. Roulet and S. Mollerach, Phys. Rep. {\bf 279}, 67 (1997).
%
\bibitem{birrell}
N.D. Birrell and P.C.W. Davies, {\em Quantum fields in curved space}
(Cambridge University Press, Cambridge, England, 1982).

%
\end{thebibliography}
\end{document}